\documentclass{llncs}

\usepackage{amssymb}
\usepackage{amsmath}
\usepackage{upgreek}
\usepackage{url}
\usepackage[colorlinks=true,urlcolor=blue,linkcolor=blue,citecolor=blue]{hyperref}

\usepackage[utf8x]{inputenc}
\usepackage{bbm}
\usepackage{graphicx}
\usepackage{listings}
\usepackage{xcolor}
\definecolor{LtGray}{rgb}{0.95,0.95,0.95}

\lstdefinelanguage{maude}{ keywords={pr, ex, inc, fmod, endfm, mod,
    is, endm, sort, sorts, subsort, op, ops, eq, rl, ceq, if, then,
    else, fi, crl, assoc, comm, ctor, id, var, vars, mb, cmb, view, endv, to, format, variant, narrowing} }
\lstnewenvironment{maude}[1][]{\lstset{language=maude,#1}}{}

\begin{document}

\title{$\uppi$: Towards a Simple Formal Semantic Framework for Compiler Construction}


\author{Christiano Braga\\\email{cbraga@ic.uff.br}}

\institute{Instituto de Computa\c{c}\~ao\\Universidade Federal Fluminense}

\maketitle

\begin{abstract}
This paper proposes $\uppi$, a formal semantic framework for compiler construction together with program validation. $\uppi$ is comprised by $\uppi$ Lib, a set of programming languages constructs inspired by Peter Mosses' Component-Based Semantics and $\uppi$ Automata, an au\-to\-mata-based formalism to describe the operational semantics of programming languages, that generalizes Gordon Plotkin's Interpreting Automata.   
\end{abstract}






\section{Introduction}\label{sec:intro}

Compiler construction is considered an intimidating discipline in Computer Science and related courses. This is perhaps captured quite graphically by the cover of the standard book on the subject, the so-called ``Dragon book" (Compilers: Principles, Techniques, and Tools~\cite{Aho:2006:CPT:1177220}), by Alfred V. Aho, Jeffrey D. Ullman and later on with Ravi Sethi and Monica S. Lam. There are ``red'', ``green'' and ``purple dragon'' editions, but the Dragon, representing how burdensome people think of the subject, is always there. 

The author has been developing and applying a formal approach, called \href{http://github.com/ChristianoBraga/BPLC}{$\uppi$}, for compiler construction, aiming at a simple technique, that relies on basic mathematics and standard Computer Science courses, that could eventually ease compiler construction and help teaching the subject. Event though preliminary attempts on its \href{http://www2.ic.uff.br/~cbraga/pmwiki/pmwiki.php/Classes/Compiladores}{pedagogical use} have been made, \emph{the main objective of this paper is to present the framework and its implementation in the Maude language.}


$\uppi$ is comprised by $\uppi$ Lib, a set of programming languages constructs inspired by Peter Mosses' \href{https://plancomps.github.io/CBS-beta}{Component-Based Semantics}~\cite{Mosses:2008:CDP:2227536.2227559} and $\uppi$ Automata, an au\-to\-mata-based formalism to describe the operational semantics of programming languages, that generalizes Gordon Plotkin's Interpreting Automata approach~\cite{plotkin}.   
To write a compiler using $\uppi$, one needs to transform the (abstract) syntax tree of a given language into a description in $\uppi$ Lib. Then, one can execute, validate or have machine code using different formal tools developed for $\uppi$ Lib, such as an interpreter, a model checker, or a code generator, implemented following the formal semantics of $\uppi$ Lib given in terms of $\uppi$ Automata. 


This paper contributes with $\uppi$, an automata-based semantic framework for formal compiler construction and its implementation in the Maude language. A Python implementation as a \href{https://nbviewer.jupyter.org/github/ChristianoBraga/BPLC/blob/master/python/pi.ipynb}{Jupyter notebook} is also underway, to explore different compilation and validation techniques.
In this paper, we will focus on $\uppi$ Automata for the \emph{dynamic semantics} of programming languages, and its Maude implementation. 
\emph{For the moment}, parsing and transformation to $\uppi$ Lib depend on the particular framework used to implement $\uppi$, Maude in this paper. Optimization is also left to an external framework, such as LLVM. In the foreseeable future we intend to cover all phases of the compiler construction process, in a formal way, based on $\uppi$ Automata.

The remainder of this paper is organized as follows. In Section~\ref{sec:rel-work} related work is discussed. Section~\ref{sec:preliminaries} recalls some preliminary material to the discussion of $\uppi$ Automata, subject of Section~\ref{sec:gia}. The $\uppi$ Automata semantics of $\uppi$ Lib is discussed in Section~\ref{sec:uppi-lib-sig}, together with its Maude implementation. 
Section~\ref{sec:conclusion} concludes this paper with the usual final remarks and indication of future work.

\section{Preliminaries}\label{sec:preliminaries}


\subsection{Transition systems, structural operational semantics and model checking}\label{sec:pre-os}

This section recalls, very briefly, just for completeness, the basic concepts of labeled and unlabeled transition systems, structural operational semantics and model checking.

A transition system~\cite{arnold1994finite} (TS) is a pair $\mathcal{T} = (S, \rightarrow)$, where $S$ denotes the set of the states of the system, $\rightarrow \subseteq S \times S$ is the transition relation. 
%
Transition systems are the standard models of structural operational semantics (SOS) descriptions.\footnote{As a matter of fact, SOS has \emph{labeled} transition systems as models with the set of labels denoting actions of the system. Labels are essential while modeling action synchronization in concurrent systems. Therefore, since we are not discussing concurrency primitives in this paper, considering the more liberal transition systems as models of SOS descriptions will not cripple the proposal of this paper.} Given an SOS description $\mathcal{M} = (G, R)$ specifying the semantics of a programming language $L$, the set $G$ defines the grammar of $L$ while relation $R$ represents the semantics of $L$ (either static or dynamic) in a syntax-directed way.   
Rule~\ref{eq:sos-congr} presents the general form of the transition rule for 
the inductive step of the evaluation of a programming language
construct $f$ in the SOS framework, where $\rho$ and $\rho'$ are environments, $\rho'$ is the result of
some computation involving $\rho$, $f$ is a programming language
construct and $t_i$ its parameters,
$\sigma, \sigma', \sigma_i, \sigma'_i$ are memory stores, with
$\sigma'$ the result of some computation of $\uplus^n_{i=1}\sigma'_i$, and
$\mathit{Cnd}$ is a predicate not involving transitions.
\begin{equation}\label{eq:sos-congr}
\frac{\rho' \vdash \bigwedge^n_{i=1} t_i, \sigma_i \Rightarrow t'_i, \sigma'_i}
     {\rho \vdash f(t_1, t_2, \ldots, t_n), \sigma
       \Rightarrow f(t'_1, t'_2, \ldots, t'_n), \sigma'} \mbox{ if } \mathit{Cnd}
\end{equation}
Typically, SOS rules have a \emph{sequent} in the conclusion of the form $\rho \vdash g , \sigma \Rightarrow g' , \sigma'$, where $g$ and $g'$ are derivations of $G$. If one looks at the conclusion as a transition of the form $(\rho, g, \sigma) \Rightarrow (\rho, g', \sigma')$ then the construction of the transition system $\mathcal{T}$ from $\mathcal{M}$ becomes straightforward with $(\rho, g, \sigma) \in S$.

Model checking~\cite{Clarke:2000:MC:332656} is an automata-based automated validation technique to solve the ``question'' 
$\mathcal{T}, s \models \varphi$, that is, does model $\mathcal{T}$, a transition system, or Kripke structure in Modal Logic~\cite{goldblatt} jargon, with initial state $s$, satisfies property $\varphi$? The standard algorithm checks if the language accepted by the intersection B\"uchi automaton (a regular $\Omega$-automaton, that is, an automaton that accepts infinite words) of $\mathcal{T}$ and $\neg\varphi$ is empty.  

\subsection{Maude}\label{sec:pre-maude}

In this section we introduce the main elements of the Maude language, our choice of programming language for this work. 

The Maude system and language~\cite{Clavel:2007:MHL:1808998} is a high-performance implementation of Rewriting Logic~\cite{meseguer92}, a formalism for the specification of concurrent systems that has been shown to be able to represent quite naturally many logical and semantic frameworks~\cite{marti-oliet-meseguer:2002}.

Maude\footnote{Maude allows for programming with different Equational Logics: Many-sorted, Order-sorted or Membership Equational Logic. In this paper, Maude programs are described using Order-sorted Equational Logic.}
is an algebraic programming language. A program in Maude is organized by modules, and every module has an initial algebra~\cite{Goguen:1996:ASI:547173} semantics. Module inclusion may occur in one of three different \emph{modes}: \texttt{including}, \texttt{extending} and \texttt{protecting}. The \texttt{including} mode is the most liberal one and imposes no constraints on the preservation of the algebra of the included module into the including one, that is, both ``junk''  and ``confusion''\footnote{Informally, when ``junk'' may be added  to an algebra but ``confusion'' may not, as in \texttt{extending} mode, it means that new terms may be included but are not identified with old ones.} may be added. Inclusion in \texttt{extending} mode may add ``junk'' but no ``confusion'', while inclusion in \texttt{protecting} mode adds no ``junk'' and no ``confusion'' to the included algebra. Module inclusion is not enforced by the Maude engine, being understood only as an indication of the intended inclusion semantics. Such declarations, however, are part of the semantics of the module hierarchy and may be important for Maude-based tools, such as a theorem prover for Maude specifications, that would have to discharge the proof obligations generated by such declarations. 

Computations in Maude are represented by rewrites according to either equations, rules or both in a given module. Functional modules may only declare equations while system modules may declare both equations and rules. Equations are assumed (that is, yield proof-obligations) to be Church-Rosser and terminating~\cite{Baader:1998:TR:280474}. Rules have to be coherent: no rewrite should be missed by alternating between the application of rules and equations. 
A (concurrent) system is specified by a rewrite system $\mathcal{R} = (\Sigma, E \cup A, R)$ where $\Sigma$ denotes its signature, $E$ the set of equations, $A$ a set of axioms, and $R$ the set of rules. The equational theory $(\Sigma, E \cup A)$ specifies the \emph{states} of the system, which are terms in the $\Sigma$-algebra modulo the set of $E$ equations and $A$ axioms, such as associativity, commutativity and identity. Combinations of such axioms give rise to different rewrite theories such that rewriting takes place \emph{modulo} such axioms. Rules $R$ specify the (possibly) non-terminating behavior, that takes place modulo the equational theory $(\Sigma, E \cup A)$. 
Another interesting feature of Maude is to support \emph{non-linear patterns} (when the same variable appears more than once in a pattern) both in equations and rules. Section~\ref{sec:gia-in-maude} exemplifies how this feature is intensively used in the Maude implementation of the $\uppi$ Automata framework.

An interesting remark regards the decision between modeling behavior as equations or rules.  One may specify (terminating) system behavior with equations. The choice between equations and rules provides an \emph{observability gauge}. In the context of a software architecture, for instance, \emph{non-observable} (terminating) actions, internal to a given component, may be specified by equations, while \emph{observable} actions, that relate components in a software architecture, may be specified as rules.  Section~\ref{sec:gia-in-maude} illustrates how this ``gauge'' is used in the Maude implementation of the $\uppi$ framework.  

A compiler can be implemented in Maude as a \emph{meta-level} application. Such a Maude application uses the so called \emph{descent functions}~\cite[Ch.11]{maude} that represent modules as terms in a \emph{universal} theory, implemented in Maude as a system module called META-LEVEL. Some of the descent functions are metaParse, metaReduce, metaRewrite and metaSearch. 
\begin{itemize}
\item Function metaParse receives a (meta-represented) module denoting a grammar, a set of quoted identifiers representing the (user) input and a quoted identifier representing the rule that should be applied to the given input qids, and returns a term in the signature of the given module. 
\item Descent function metaReduce receives a (meta-represented) module and a (meta-represented) term and returns the (meta-represented) canonical form of the given term by the exhaustive application of the (Church-Rosser and terminating) \emph{equations}, only, of the given module.  An interesting example of metaReduce is the invocation of the model checker at the meta-level: (i) first, module MODEL-CHECKER must be included in a module that also includes the Maude description of the system to be analyzed, and (ii) one may invoke metaReduce of a meta-representation of a term that is a call to function modelCheck, with appropriate parameters, defined in module MODEL-CHECKER. 
\item Finally, function metaRewrite simplifies, in a certain number of steps, a given term according to both equations and rules (assumed coherent, that is, no term is missed by the alternate application of equations and rules) of the given module. The descent function metaSearch looks for a term that matches a given \emph{pattern}, from a given term, according to a choice of rewrite relation from $\Rightarrow^*$, $\Rightarrow^+$, $\Rightarrow^!$, denoting the reflexive-transitive closure of the rewrite relation, the transitive closure of the rewrite relation or the rewrite relation that produces only canonical forms.
\end{itemize}


\section{$\uppi$ Automata}\label{sec:gia}

\subsection{Interpreting automata}

In~\cite{plotkin}, Plotkin defines the concept of Interpreting Automata as finite-state Transition Systems as a semantic framework for the operational semantics of programming languages. Interpreting Automata are now recalled from the perspective of Automata Theory.

Let $\mathcal{L}$ be a programming language accepted by a Context Free Grammar (CFG) $G = (V, T, P, S)$ defined in the standard way where $V$ is the finite set of variables (or non-terminals), $T$ is the set of terminals, $P \subseteq V \times (V \cup T)^*$ and $S \not\in V$ is the start symbol of $G$.
An Interpreting Automaton for $\mathcal{L}$ is a tuple $\mathcal{I} = (\Sigma, \Gamma, \rightarrow, \gamma_0, F)$ where $\Sigma = T$, $\Gamma$ is the set of configurations, $\rightarrow \subseteq \Gamma \times \Gamma$ is the transition relation, $\gamma_0 \in \Gamma$ is initial configuration, and $F$ the finite set of final configurations. Configurations in \(\Gamma\) are triples of the form
\(
\Gamma = \mathit{Value~Stack} \times \mathit{Memory} \times \mathit{Control~Stack},
\)
where $\mathit{Value~Stack} = (L(G))^*$ with $L(G)$ the language generated by $G$,\footnote{There are some situations where one may need to push not only computed values but \emph{code} as well into the value stack. One such situation is when a \emph{loop} is being evaluated and both the loop's test and body are pushed in order to ``reconstruct'' the loop for the next iteration.} the set
\(\mathit{Memory}\) is a finite map \(\mathit{Var}
\to_{\mathit{fin}} \mathit{Storable}\) with $\mathit{Var} \in V$ and $\mathit{Storable} \subseteq T^*$, and the elements of the
$\mathit{Control~Stack} = (L(G) \cup \mathit{KW})^*$, where $\mathit{KW}$ is the set of keywords of $\mathcal{L}$. 
A computation in $\mathcal{I}$ is defined as $\rightarrow^*$, the reflexive-transitive closure of the transition relation.

As an example, let us consider the CFG of a programming language $\mathcal{L}$ with arithmetic expressions,
Boolean expressions and commands. 
\[
\begin{array}{rcl}
\mathit{Prog} & ::= &  \mathit{ComSeq} \\
\mathit{ComSeq} & ::= & \mathtt{nop} \mid \mathit{Com} \mid \mathit{Com} ~\mathtt{;}~ \mathit{ComSeq}   \\
\mathit{Com} & ::= & \mathit{Var} ~\mathtt{:=}~ \mathit{Exp} \mid \\
		     &  & \mathtt{if}~ \mathit{BExp}~ \mathit{ComSeq}~ \mathit{ComSeq} \mid \\
		     &  & \mathtt{while}~ \mathit{BExp}~ \mathit{ComSeq} \\
\mathit{Exp} & ::= & \mathit{BExp} \mid \mathit{AExp} \\
\mathit{BExp} & ::= & \mathit{Exp} ~\mathit{BOP}~ \mathit{Exp} \\
\mathit{BOP} & ::= & \mathtt{=} \mid \mathtt{or} \mid \mathtt{∼} \\  
\mathit{AExp} & ::= & \mathit{AExp} ~\mathit{AOP}~ \mathit{AExp} \\
\mathit{AOP} & ::= & \mathtt{+} \mid \mathtt{-} \mid \mathtt{*} 
\end{array}
\]

The values in the $\mathit{Value~Stack}$ are
elements of the set 
$\mathbbm{T} \cup \mathbbm{N} \cup \mathit{Var} \cup \mathit{BExp} \cup \mathit{Com},$ 
where \(\mathbbm{T}\) is the
set of Boolean values, \(\mathbbm{N}\) is the set of natural numbers,
with \(\mathit{Var}\) the set of variables, \(\mathit{BExp}\) the set
of Boolean expressions, and \(\mathit{Com}\) the set of commands of $\mathcal{L}$.
The $\mathit{Control~Stack}\) is defined as the set 
$(\mathit{Com} \cup \mathit{BExp} \cup \mathit{AExp} \cup \mathit{KW})^*$, where
\(\mathit{AExp}\) is the set of arithmetic expressions and 
$\mathit{KW}= \{\mathtt{+}, \mathtt{−}, \mathtt{∗}, \mathtt{=}, \mathtt{or}, \mathtt{∼},\mathtt{:=},\mathtt{if}, \mathtt{while}, \mathtt{;}\}$.

Informally, the computations of an Interpreting Automaton mimic the behavior of a calculator
in Łukasiewicz postfix notation, also known as reverse Polish notation. A typical computation of an Interpreting Automaton
\emph{interprets} a statement $c(p_1, p_2, \ldots, p_n) \in L(G)$ on the top of $\mathit{Control~Stack}$ \(C\) of a
configuration \(\gamma = (S, M, C)\), by unfolding its subtrees $p_i \in L(G)$ and $c \in \mathit{KW}$ that are then pushed back into $C$, and possibly updating the $\mathit{Value~Stack}$
\(S\) with intermediary results of the interpretation of the $c(p_1, p_2, $\ldots$, p_n)$, and the $\mathit{Memory}$, should $c(p_1, p_2, $\ldots$, p_n) \in L(\mathit{Com})$.

For the transition relation of $\mathcal{I}$, let us consider the rules for
arithmetic sum expressions. 
\begin{eqnarray}
\label{eq:val}\langle S, M, n~ C \rangle & \Rightarrow & \langle n ~ S, M, C \rangle \\
\label{eq:add1}\langle S, M, (e_1 \mathtt{+} e_2) ~ C \rangle & \Rightarrow &
\label{eq:add2} \langle S, M, e_1 ~ e_2 ~ \mathtt{+}~ C \rangle \\
\langle n ~ m ~ S, M, \mathtt{+}~ C \rangle & \Rightarrow & \langle (n + m)~ S, M, C \rangle
\end{eqnarray}
where \(e_i\) are metavariables for arithmetic expressions, and \(n, m
\in \mathbbm{N}\).  Rule~\ref{eq:add1} specifies that when the
arithmetic expression \(e_1 \mathtt{+} e_2\) is on top of the control
stack \(C\), then its operands should be pushed to \(C\) and then the
operator \texttt{+}. Operands \(e_1\) and \(e_2\) will be recursively
evaluated, as a computation is the reflexive-transitive closure of
relation \(\rightarrow\), leading to a configuration with an element in $\mathbbm{T} \cup \mathbbm{N}$ left on
top of the value stack $S$, as specified by
Rule~\ref{eq:val}. Finally, when $\mathtt{+}$ is on top of the control
stack $C$, and there are two natural numbers on top of $S$, they are
popped, added and pushed back to the top of $S$.

Finally, there is one quite interesting characteristic of Interpreting
Automata:
\emph{transitions do not appear in the conditions of the rules}, a
characteristic that can be quite desirable from a proof theoretic
standpoint, in particular in the context of term rewriting systems (see Section~\ref{sec:gia-and-trs}), as
pointed out by Viry~\cite{zbMATH01440294} and later by Ro\c{s}u in~\cite{10.1007/978-3-540-31959-7_13}, for instance. 
As opposed to transition rules that admit transitions in its premises, as in the Structural Operational Semantics (SOS) framework, for instance, also defined in~\cite{plotkin}, Interpreting Automata evaluation uses the control stack to push the evaluation context, so to speak, to the configuration. Unconditional rewriting has also very desirable computational consequences, in particular in Maude, regarding executability and performance. Model checking, for instance, does not consider transitions (rewrites) in the conditions of rules. Also, narrowing does not work, for the time being, on conditional rules. Regarding performance, the combination of a proper use of equations instead of rules together with unconditional rules provides an effective search mechanism. The use of equations shortens the state space and unconditional rules do not create ``scratch pad''  
rewrites performing only forward rewriting.

As an example, let us recall Rule~\ref{eq:sos-congr}, the general form of the rule for 
the recursive step of the evaluation of a programming language
construct $f$ in the SOS framework, 
\begin{equation}
\frac{\rho' \vdash \bigwedge^n_{i=1} t_i, \sigma_i \Rightarrow t'_i, \sigma'_i}
     {\rho \vdash f(t_1, t_2, \ldots, t_n), \sigma
       \Rightarrow f(t'_1, t'_2, \ldots, t'_n), \sigma'} \mbox{ if } \mathit{Cnd} \nonumber
\end{equation}
where $\rho$ and $\rho'$ are environments, $\rho'$ is the result of
some computation involving $\rho$, $f$ is a programming language
construct and $t_i$ its parameters,
$\sigma, \sigma', \sigma_i, \sigma'_i$ are memory stores, with
$\sigma'$ the result of some computation of $\sigma'_i$, and
$\mathit{Cnd}$ is a predicate not involving transitions.

The Interpreting Automata rule for Rule~\ref{eq:sos-congr} is as follows,
\begin{equation}\label{eq:ia-congr}
\{f(t_1, t_2, \ldots, t_n)~ C, \rho, \sigma\} \Rightarrow
\{t_1 t_2 \ldots t_n f ~ C, \rho, \sigma\} \mbox{ if } \mathit{Cnd},
\end{equation}
where $C$ is the control stack. Note that recursion will take care of
evaluating a $t_i$ when it is on top of the control stack, so there is
no need to explicitly require transitions of the form $t_i \Rightarrow
t'_i$ as premises or conditions to Rule~\ref{eq:ia-congr}. Copies of
the environment (such as $\rho'$ in Rule~\ref{eq:sos-congr}) and
side-effects are \emph{naturally} calculated during the computation
process by the application of the appropriate rule for the term on top
of the control stack.

\subsection{$\uppi$ Automata}

$\uppi$ Automata are Interpreting Automata
whose \emph{configurations are sets of semantic components} that include, at least, 
a $\mathit{Value~Stack}$, a $\mathit{Memory}$ and a $\mathit{Control~Stack}$.
Plotkin's stacks and memory in
Interpreting Automata (or environment and stores of Structural
Operational Semantics) are generalized to the concept of \emph{semantic
  component}, as proposed by Peter Mosses in the Modular
SOS approach to the formal semantics of
programming languages.

Formally, a $\uppi$-automaton is an Interpreting Automaton where, given an abstract finite pre-order \emph{Sem}, for semantics components, its configurations are defined by $\Gamma = \uplus^n_{i=1} \mathit{Sem}$, with $n
\in \mathbbm{N}$, $\uplus$ denoting the disjoint union operation of
$n$ semantic components, with $\mathit{Value~Stack}$, $\mathit{Memory}$ and $\mathit{Control~Stack}$ subsets of $\mathit{Sem}$.
%


The semantic rules for arithmetic sum in $\uppi$ Automata look
very similar to the ones from Interpreting Automata. 
\begin{eqnarray}
\label{eq:gia-val} \{ S, M, n~ C, \ldots \} & \Rightarrow & \{ n ~ S, M, C, \ldots \} \\
\label{eq:gia-add1}\{ S, M, (e_1 \mathtt{+}~ e_2) ~ C, \ldots \} & \Rightarrow &
\label{eq:gia-add2}\{ S, M, e_1 ~ e_2 ~ \mathtt{+}~ C, \ldots \} \\
\{ n ~ m ~ S, M, \mathtt{+}~ C , \ldots \} & \Rightarrow & \{ (n + m)~
S, M, C, \ldots \}
\end{eqnarray}
The ellipsis ``$\ldots$'' \footnote{This notation is similar to the one defined by
  Chalub and Mosses in the Modular SOS Description Formalism, which is
  implemented in the Maude MSOS Tool~\cite{wrla06}.} are adopted as notation for
``don't care'' semantic components, that is, those components that are
not relevant for the specification of the semantics of a particular
language construct.

The point is that if one wants to extend one's Interpreting Automata specification with new
semantic components, say disjointly uniting an output component (representing standard output in the C language, for instance), understood as a
sequence of values, to the already existing disjoint set of
environments and stores, would require a reformulation of the existing
specification. For instance, the specification for arithmetic sum in Interpreting Automata 
would require such reformulation while in $\uppi$ Automata would not. 
The rules in the latter have
the ``don't care'' variable that matches any, or no component at all,
that may be together with $S$, $M$ and $C$. 
Semantic component
composition is \emph{monotonic}, as the addition of new semantic
components does not affect the transition relation, that is, $x
\Rightarrow y ~\mbox{implies}~ g(x) \Rightarrow g(y)$, where $x, y \in \Gamma$
and $g$ is a function that adds a new semantic component to $\Gamma$.
%

\subsection{$\uppi$ Automata and Term Rewriting}\label{sec:gia-and-trs}

A $\uppi$-automaton $\mathcal{I} = (\Sigma, \Gamma, \rightarrow, \gamma_0, F)$ 
can be seen as an unlabeled Transition System $\mathcal{I} = (\Gamma, \rightarrow)$ and therefore  
as a Term Rewriting
System~\cite{Baader:1998:TR:280474} when the latter is understood as
$\mathcal{T} = (A, \longrightarrow)$ where $A$ is a set and $\longrightarrow$ a
reduction relation on $A$. Clearly, the set of configurations $\Gamma$ is $A$
and the transition relation of the Interpreting Automata is the
reduction relation of the Term Rewriting System.

There is an interesting point on the relation between the
semantics of a programming language construct, specified by a
$\uppi$ Automata, and the \emph{properties} that one
may require from the reduction relation of the associated Term
Rewriting System (TRS). Let us first recall two basic properties of a reduction relation
from~\cite[Def.2.1.3]{Baader:1998:TR:280474},
\begin{itemize}
\item Church-Rosser: $x \stackrel{*}{\longleftrightarrow} y \implies x
  \downarrow y$,
\item termination: there is no infinite reduction $a_0 \rightarrow a_1
  \rightarrow \ldots$.
\end{itemize}
where $x, y \in A$, $\stackrel{*}{\longleftrightarrow}$ denotes the
reflexive-transitive-symmetric closure of $\longrightarrow$, and
$x \downarrow y$ denotes that $x$ and $y$ are joinable, that is,
$\exists z, x \stackrel{*}{\longrightarrow} z
\stackrel{*}\longleftarrow y$.  In rewriting modulo equational
theories~\cite[Ch. 11]{Baader:1998:TR:280474}, otherwise
non-terminating systems become terminating when an algebraic property,
such as commutativity, is incorporated into the rewriting process. Given a
TRS $(A, \longrightarrow)$, let $E$ be a set with
the identities induced by a given property, such as commutativity, and
$R$ the remaining identities induced by $\longrightarrow$. Rewriting
then occurs on equivalence classes of terms, giving rise to a new
relation, $\longrightarrow_{R/E}$, defined as follows:
$$
[s]_{\approx E} \longrightarrow_{R/E} [t]_{\approx E} \Leftrightarrow
\exists s', t'. s \approx_E s' \longrightarrow_R t' \approx_E t.
$$


Moving back to $\uppi$ Automata, the
semantics of a programming language construct $c(p_1, \ldots, p_n)$ is
\emph{functional}, where $c$ is the construct and $p_i$ its
parameters, when given any configuration
$\gamma = \{ c(p_1, \ldots, p_n) ~C, \ldots\}$, there exists a single
$\gamma'$ such that $\gamma \Rightarrow^* \gamma'$ and the computation is finite. The semantics of a
programming language construct $c(p_1, \ldots, p_n)$ is
\emph{relational} when given any configuration
$\gamma = \{ c(p_1, \ldots, p_n)~ C, \ldots\}$, where
$C \in \mathit{Control~Stack}$, the computations starting in $\gamma$ may lead to different $\gamma'_i$ and 
may not terminate.

Therefore, if the semantics of a programming language construct is
functional, one must require the associated reduction relation to be
Church-Rosser and terminating. No constraints are imposed to the
reduction relation when the semantics is relational.

As an illustration, according to this definition, the semantics of
addition is \emph{functional} but an undefined loop (such as a while
command) semantics is \emph{relational} as its execution may not
terminate.

In order to support the specification of monotonic rules in a modular
way, one last thing is required from the TRS
associated with a $\uppi$ Automata: rewriting modulo
associativity, idempotence and commutativity. In other words, \emph{set}-rewriting takes place, not simply term rewriting, while
representing $\uppi$ Automata as TRS, as each rule rewrites a set of semantic components.


\subsubsection{$\uppi$ Automata in Maude}\label{sec:gia-in-maude}

Maude parameterized programming capabilities are used to implement 
$\uppi$ Automata. The main datatype of $\uppi$ Automata is Generalized SMC (GSMC in Listing~\ref{lst:gsmc-maude}), a disjoint union set of semantic components. 
The trivial view SemComp maps terms of sort Elt to terms of sort SemComp.   
Module GSMC then imports module SET parameterized by
view SemComp, of semantic components,
implemented in Maude by functional module GSMC-SORTS. 
A configuration of a $\uppi$-automaton is declared with 
constructor \texttt{<\_> : Set{SemComp} -> Conf} that gives rise to terms such as 
\texttt{< $c_1$, $c_2$ >} where \texttt{$c_i$} is a semantic component.

\begin{maude}[caption=Generalized SMC in Maude,label=lst:gsmc-maude]
fmod SEMANTIC-COMPONENTS is sorts SemComp . endfm
view SemComp from TRIV to SEMANTIC-COMPONENTS is sort Elt to SemComp . endv
fmod GSMC is ex VALUE-STACK . ex MEMORY . ex CONTROL-STACK . ex ENV . 
    ex SET{SemComp} * (op empty to noSemComp) .
    sorts Attrib Conf EnvAttrib StoreAttrib ControlAttrib ValueAttrib .
    subsort EnvAttrib StoreAttrib ControlAttrib ValueAttrib < Attrib  .
    op <_> : Set{SemComp} -> Conf [format(c! c! c! o)] . 
    op env : -> EnvAttrib .     --- Semantic components
    op sto : -> StoreAttrib .
    op cnt : -> ControlAttrib .
    op val : -> ValueAttrib .
    op _:_ : EnvAttrib Env -> SemComp [ctor format(c! b! o o)] .
    op _:_ : StoreAttrib Store -> SemComp [ctor format(r! b! o o)] .
    op _:_ : ControlAttrib ControlStack -> SemComp [ctor format(c! b! o o)] .
    op _:_ : ValueAttrib ValueStack -> SemComp [ctor format(c! b! o o)] .
endfm
\end{maude}

%
%
%

Recall that the elements of the disjoint union $\biguplus^n_{i=1} \mathit{Sem}$ are ordered pairs $(s, i)$ such that $i$ serves as an index indicating which semantic component $s$ came from. This is exemplified in Maude with the memory store component. 
The constructor operator \texttt{sto} functions as the index for the memory store component and the constructor operator \texttt{\_:\_} to represent ordered pairs $(s, i)$ where $s$ is the memory store.  

Now, for the transition rules, they are represented either by equations or rules, depending on the semantic character of the programming language construct being formalized. In the case of arithmetic expressions 
their character is functional and therefore are implemented as equations in Maude. For sum, in equation \texttt{add-exp1}\footnote{Keyword \texttt{variant} is an attribute for equations and means that the given equation should be used in the variant unification process. Due to space constraints, this feature is not discussed in this paper. The keyword is left in the code snippet to present the actual executable code for the tool.}  first operands \texttt{E1:Exp} and \texttt{E2:Exp} are unfolded, and then pushed back to the control stack \texttt{C}, together with \texttt{ADD}, an element of set $\mathit{KW}$. (Recall that $\mathit{Control~Stack} ~=~ (\mathit{Com} \cup \mathit{BExp} \cup \mathit{AExp} \cup \mathit{KW})^*$.) Equation \texttt{add-exp2} implements the case where both \texttt{E1:Exp} and \texttt{E2:Exp} have been both evaluated and their associated (Rational) value (in this implementation) was pushed to the value stack. When \texttt{ADD} is on top of the control stack then the two top-most values in the value stack are added. (Note that \texttt{+} symbol in \texttt{add-exp2} denotes sum in the Rationals whereas in  \texttt{add-exp2} is the symbol for sum in language $\mathcal{L}$.)
\begin{maude}
eq [add-exp1] : < cnt : (E1:Exp + E2:Exp) C:ControlStack), ... >  =
               < cnt : (E1:Exp E2:Exp ADD C:ControlStack), ... > [variant] .
eq [add-exp2] : < cnt : (ADD C:ControlStack), 
                 val : (val(R1:Rat) val(R2:Rat) SK:ValueStack), ... >  =
               < cnt : C:ControlStack, 
                 val : (val(R1:Rat + R2:Rat) SK:ValueStack), ... > [variant] .
\end{maude}

\subsection{Model checking $\uppi$ Automata}\label{sec:mc-gia}

Model checking (e.g.~\cite{Clarke:2000:MC:332656}) is perhaps the most popular formal method for the validation of concurrent systems. The fact that it is an \emph{automata-based automated validation technique} makes it a nice candidate to join a simple framework for teaching language construction that also aims at validation, such as the one proposed in this paper.


This section recalls the syntax and semantics for (a subset of) Linear Temporal Logic, one of the Modal Logics used in model checking, and discusses how to use this technique to validate $\uppi$ Automata, only the necessary to follow Section~\ref{sec:imp}.

The syntax of Linear Temporal Logic is given by the following grammar 
$$\begin{array}{c}
\phi ::= \top ~|~ \bot ~|~ p ~|~ \neg(\phi) ~|~ (\phi \land \phi) ~|~ (\phi \lor \phi) ~|~ (\phi \to \phi) ~|~ 
(\Diamond \phi) ~|~ (\Box \phi) 
\end{array}$$
where connectives $\Diamond$, $\Box$
are called \emph{temporal modalities}. They denote ``Future state'' and ``Globally (all future states)''. There is a precedence among them given by: first unary modalities, in the following order $\neg$, $\Diamond$ and $\Box$, then binary modalities, in the following order, $\land, \lor$ and $\to$. 

The standard models for Modal Logics (e.g.~\cite{goldblatt}) are Kripke structures, triples $\mathcal{K} = (W, R, L)$ where $W$ is a set of worlds, $R \subseteq W \times W$ is the world accessibility relation and $L : W \to 2^{\mathit{AP}}$ is the labeling function that associates to a world a set of atomic propositions that hold in the given world. Depending on the modalities (or operators in the logic) and the properties of $R$, different Modal Logics arise such as Linear Temporal Logic.
A \emph{path} in a Kripke structure $\mathcal{K}$ represents a possible (infinite) scenario (or computation) of a system in terms of its states. The path $\tau = s_1 \to s_2 \to \ldots$ is an example. A \emph{suffix} of $\tau$ denoted $\tau^i$ is a sequence of states starting in $i$-th state.
Let $\mathcal{K} = (W, R, L)$ be a Kripke structure and $\tau = s_1 \to \ldots$ a path in $\mathcal{K}$. Satisfaction of an LTL formula $\phi$ in a path $\tau$, denoted $\tau \models \varphi$ is defined as follows, 
$$\begin{array}{l}
\tau \models \top, \qquad \tau \not\models \bot, \qquad \tau \models p ~\mathit{iff}~ p \in L(s_1), \quad \tau \models \neg\phi ~\mathit{iff}~ \tau \not\models \phi, \\
\tau \models \phi_1 \land \phi_2 ~\mbox{iff}~ \tau \models \phi_1 ~\mbox{and}~ \tau \models \phi_2, \\
\tau \models \phi_1 \lor \phi_2 ~\mbox{iff}~ \tau \models \phi_1 ~\mbox{or}~ \tau \models \phi_2, \\
\tau \models \phi_1 \to \phi_2 ~\mbox{iff}~ \tau \models \phi_2 ~\mbox{whenever}~ \tau \models \phi_1, \\
\tau \models \Box \phi ~\mbox{iff}~ \mbox{for all}~ i \ge 1, \tau^i \models \phi, \\
\tau \models \Diamond \phi ~\mbox{iff}~ \mbox{there is some}~ i \ge 1, \tau^i \models \phi. \\
\end{array}$$

A $\uppi$ Automata, when understood as a Transition System, is also a \emph{frame}, that is, $\mathcal{F} = (W, R)$, where $W$ is the set of worlds and $R$ the accessibility relation. A Kripke structure is defined from a frame representing a $\uppi$ Automata by declaring the labeling function with the following state proposition scheme:
\begin{eqnarray}
\forall \sigma \in \mathit{Memory}, v \in \mathit{Index}(\sigma), r \in \mathit{Storable}, \nonumber\\
\langle \sigma, \ldots \rangle \models p_v(r) =_{\mathit{def}} (\sigma(v) = r),
\end{eqnarray}
meaning that for every variable $v$ in the index of the memory store component (which is a necessary semantic component) there exists a unary proposition $p_v$ that holds in every state where $v$ is bound to $p_v$'s parameter in the memory store.  A \emph{poetic license} is taken here and $\uppi$ Automata, from now on, refers to the pair composed by a $\uppi$ Automata and its state propositions.
As an illustrative specification, used in Section~\ref{sec:imp}, the LTL formula $\Box \neg[p_1(\mathit{crit}) \land p_2(\mathit{crit})]$ specifies safety (``nothing bad happens''), in this case both $p_1$ and $p_2$ in the critical section, when $p_i$ are state proposition formulae denoting the states of two processes and $\mathit{crit}$ is a constant denoting that a given process is in the critical section, and formula $\Box[p_1(\mathit{try}) \to \Diamond (p_1(\mathit{crit}))]$ specifies liveness (``something good eventually happens''),  by stating that if a process, $p_1$ in this case, tries to enter the critical section it will eventually do so.

\section{$\uppi$ Lib: Basic Programming Language Constructs}\label{sec:uppi-lib}

$\uppi$ Lib is a subset of Constructive MSOS~\cite{Mosses:2004:FCF}, as implemented in~\cite[Ch. 6]{msc-chalub}.
In Section~\ref{sec:uppi-lib-sig}, $\uppi$ Lib constructions are presented, their $\uppi$-au\-to\-ma\-ta semantics is discussed in Section~\ref{sec:uppi-lib-gia} and a simple compiler for the \textsc{Imp} language in Maude, using $\uppi$ Lib, is described in Section~\ref{sec:imp}.

\subsection{$\uppi$ Lib signature}\label{sec:uppi-lib-sig}

The signature of $\uppi$ Lib is organized in five parts, and implemented in four different modules in Maude: (i) Expressions, that include basic values (such as Rational numbers and Boolean values), identifiers, arithmetic and Boolean operations, (ii) Commands, statements that produce side effects to the memory store, (iii) Declarations, which are statements that construct the constant environment, (iv) output and (v) abnormal termination.

Due to space constraints, only the $\uppi$ Lib signature for arithmetic expressions is discussed. 
The remaining declarations follow a similar pattern. First, it includes modules QID, RAT, and GSMC, for quoted identifiers, rational numbers and Generalized SMC machines, respectively.  Modules QID and RAT are part of the Maude standard prelude while GMSC was defined in Listing~\ref{lst:gsmc-maude}. Next, module EXP declares sorts Exp, BExp and AExp, for (general) expressions, Boolean expressions and arithmetic expressions. Identifiers are subsorts of both Boolean expressions and arithmetic expressions, which are in turn subsorts of expressions. The latter are included in Control. Operator idn constructs Identifiers from Maude built-in quoted identifiers. Arithmetic and Boolean operations alike are declared as Maude operators, and so are elements of set $\mathit{KW}$.    
\begin{maude}
fmod EXP is pr QID . pr RAT . pr GSMC .
    sorts Exp BExp AExp . subsort Id < BExp AExp < Exp < Control .
    op idn : Qid -> Id [ctor format(!g o)] . --- Identifiers
    op rat : Rat -> AExp [ctor format(!g o)] .
    op add : AExp AExp -> AExp [format(! o)] .
    op sub : AExp AExp -> AExp [format(! o)] . --- Arithmetic
    op mul : AExp AExp -> AExp [format(! o)] .
    op div : AExp AExp -> AExp [format(! o)] .
    ops ADD SUB MUL DIV : -> Control [ctor] . $\quad \ldots$
 endfm
 \end{maude}

\subsection{$\uppi$ Automata transitions for $\uppi$ Lib dynamic semantics in Maude}\label{sec:uppi-lib-gia}

Again, due to space constraints, the transition relation is not discussed for the complete $\uppi$ Lib signature. 
Transitions for loop evaluation have been chosen to illustrate $\uppi$ Automata transitions for $\uppi$ Lib.


The semantics of the loop construction in module CMD is implemented in terms of equations and a rule in Maude. 
The first equation (i) pushes the loop body into the control stack, (ii) pushes the loop test into the control stack and pushes the whole loop into the value stack. These steps are of functional character, that is, they are Church-Rosser and terminating therefore satisfying the requirements to be implemented by an equation in Maude. The execution of the body of the loop, however, may not terminate as there could be a nested loop, for instance, that does not terminate its execution. For that reason it is implemented as a rule in Maude. 
\begin{maude}
eq [loop] : 
       < cnt : loop(E:Exp, K:Cmd) C:ControlStack, val : V:ValueStack, ... >  = 
       < cnt : E:Exp LOOP C:ControlStack, val : val(loop(E:Exp, K:Cmd)) V:ValueStack, ... > 
[variant] .
rl [loop] :  
      < cnt : LOOP C:ControlStack, val : val(true) val(loop(E:Exp, K:Cmd)) V:ValueStack, ... >  =>  
      < cnt : K:Cmd loop(E:Exp, K:Cmd) C:ControlStack, val : V:ValueStack, ... >  [narrowing] .
eq [loop] : 
      < cnt : LOOP C:ControlStack, val : val(false) val(loop(E:Exp, K:Cmd))  V:ValueStack, ... >  =  
      < cnt : C:ControlStack, val : V:ValueStack, ... > [variant] .
\end{maude}

\subsection{A $\uppi$ compiler for \textsc{Imp} in Maude}\label{sec:imp}

In this Section, the use of $\uppi$ Lib is illustrated by a compiler for a simple (and yet Turing-complete) imperative language called \textsc{Imp}. 
The current implementation of $\uppi$ Lib in Maude supports execution by rewriting, symbolic execution by narrowing and LTL model-checking. These are the tools that are ``lifted'' from Maude to \textsc{Imp}.

A $\uppi$ compiler for a language $\textsc{L}$, such as \textsc{Imp}, defined as a denotation of $\uppi$ Lib constructions, has the following main components:
(i) A read-eval-loop function (or command-line interface) that invokes different meta-functions depending on the given command. For example, a load command invokes the parser, exec invokes metaRewrite and mc invokes metaReduce with the model checker.
(ii) A parser for $\textsc{L}$, which is essentially a meta-function that given a list of qids returns a meta-term according to a given grammar, specified as a functional module;
(iii) A transformer from $\textsc{L}$ to $\uppi$ Lib, a meta-function that given a term in the data-type of the source language, produces a term on the data-type of the target language;
(iv) A pretty-printer from $\uppi$ Lib to $\textsc{L}$, a meta-function that given a term in the data-type of the target language produces a list of qids. Each component is discussed next, but pretty-printing, due to space constraints.







\paragraph{\textsc{Imp}'s command-line interface}

In Listing~\ref{lst:imp-cli} we describe an excerpt of the implementation for \textsc{Imp}'s command-line interface, detailing only the module inclusions, sort declarations for the command-line state and rules for loading an \textsc{Imp} program. A full description is not possible due to space constraints. However, the pattern explained in this excerpt is the same for every command: a qidlist denotes the input, a different meta-function is called on them, depending on the input, that updates or not the state of \textsc{Imp}'s command-line interface and or the output of the system as whole, with a message to the end user.  
%
Operation \texttt{op <\_;\_;\_> : MetaIMPModule Dec? QidList -> IMPState} represents the state of the command-line interface, which is a triple comprised by (i) the meta-representation of an \textsc{Imp} module, (ii) the $\uppi$ Lib representation of the \textsc{Imp} module in the first projection, and (iii) a qid list denoting the message from the last processed command. Rule labeled \texttt{in} is responsible for processing the input of an \textsc{Imp} module, therefore, in this case, the first projection of the System term contains a list of qids representing an \textsc{Imp} module. Should the parsing process be successful, variable \texttt{T:ResultPair?} will be bound to a pair whose first projection is the term resulting from  parsing and in the second projection its sort, or a unary function, of sort \texttt{ResultPair?}, denoting that the parsing process was not properly carried on, with a parameter representing the qid where the parsing process failed. With a successful parsing, the following term becomes now the state of the system
\begin{maude}
< getTerm(T:ResultPair?) ; compileMod(getTerm(T:ResultPair?)) ; 
   'IMP: '\b 'Module Q:Qid 'loaded. '\o  >
\end{maude}
where \texttt{getTerm(T:ResultPair?)} denotes the meta-term, according to \textsc{Imp}'s grammar, representing the input module, the input module in $\uppi$ Lib is denoted by \texttt{compileMod(getTerm(T:ResultPair?))}, in the second projection, and the third component of the \texttt{IMPState} triple is a qidlist represents a message to the user know informing that the module was properly loaded.
\begin{maude}[caption=\textsc{Imp}'s command-line interface in Maude, label=lst:imp-cli]
mod IMP-INTERFACE is 
 $\ldots$
 op <_;_;_> : MetaIMPModule Dec? QidList -> IMPState .
 vars QIL QIL' QIL'' QIL1 QIL2 : QidList .
 --- Loading a module.
 crl [in] : ['module Q:Qid QIL, < M:MetaIMPModule ; D:Dec? ; QIL' >, QIL''] => 
             if (T:ResultPair? :: ResultPair) then 
			              [nil, < $\fbox{\textit{getTerm(T:ResultPair?)}}$ ;
                          $\fbox{\textit{compileMod(getTerm(T:ResultPair?))}}$ ; 
                           'IMP: '\b 'Module Q:Qid 'loaded. '\o  >, QIL'']
             else [nil, < noModule ; noDec ; nil >, 
                      printParseError('module Q:Qid QIL, T:ResultPair?)]
             fi
 if T:ResultPair? :=
    $\fbox{\textit{metaParse(upModule('IMP-GRAMMAR, false), 'module Q:Qid QIL, 'ModuleDecl)}}$ .
 $\ldots$
endm
\end{maude}

\paragraph{\textsc{Imp} parser}

To write a parser in Maude one has to first define the grammar of the language as a Maude functional module. The IMP-GRAMMAR module 
is the first argument in the function call to metaParse in Rule \texttt{in} of module IMP-INTERFACE in Listing~\ref{lst:imp-cli}, that implements \textsc{Imp}'s read-eval-loop. Essentially, variables (or non-terminals) are represented by sorts, terminals by constants and grammar rules are represented by operations. 
Module IMP-GRAMMAR encodes an excerpt of the \textsc{Imp} grammar, only enough to discuss the main elements of the grammar representation: we exemplify it with sum expressions
Sort ExpressionDecl, for instance, specifies both arithmetic and Boolean expressions. A grammar rule relating two grammar variables is represented by a subsort declaration. Therefore, PredicateDecl, the sort for Boolean expressions, is a subsort of ExpressionDecl.  Attributes \texttt{prec} and \texttt{gather} are declared for disambiguation. 
\begin{maude}
fmod IMP-GRAMMAR is pr TOKEN . pr RAT .
 inc PREDICATE-DECL . inc COMMAND-DECL .
 sorts VariablesDecl ConstantsDecl OperationsDecl ProcDeclList
       ProcDecl FormalsDecl BlockCommandDecl ExpressionDecl
       InitDecl InitDeclList InitDecls ClausesDecl
       ModuleDecl Expression .
 subsort InitDecl < InitDeclList .
 subsort VariablesDecl ConstantsDecl ProcDeclList InitDecls < ClausesDecl .
 subsort BlockCommandDecl < CommandDecl .
 subsort ProcDecl < ProcDeclList .
 subsort PredicateDecl < ExpressionDecl .
 op _+_ : Token Token -> ExpressionDecl [gather(e E) prec 15] . --- Arithmetic expressions
 op _+_ : Token ExpressionDecl -> ExpressionDecl [gather(e E) prec 15] .
 op _+_ : ExpressionDecl Token -> ExpressionDecl [gather(e E) prec 15] .
 op _+_ : ExpressionDecl ExpressionDecl -> ExpressionDecl [gather(e E) prec 15] .
$\ldots$
endfm
\end{maude}
%

\paragraph{\textsc{Imp} to $\uppi$ Lib transformer}

Compilation from \textsc{Imp} to $\uppi$ Lib is quite trivial as there exists a one-to-one correspondence between \textsc{Imp} constructions and $\uppi$ Lib. \footnote{This is not the case for variable and constant declarations that require initializations to be mapped to a $\uppi$ Lib ref declaration, a simple exercise but useful stimulate non bijectional mappings between the source language and $\uppi$ Lib. A similar situation arises when compiling \textsc{Imp} code to Python, as variable declarations are local by default, requiring the \texttt{global} modifier otherwise. This also not be the case for other programming languages, such as the denotation of object-oriented constructions~\cite{msc-chalub}.} Essentially, an \textsc{Imp} module gives rise to a $\uppi$ Lib dec. \textsc{Imp} var and const are declarations and so is a proc declaration that gives rise to a prc declaration in $\uppi$ Lib. 
%
%
The compilation from \textsc{Imp} to $\uppi$ Lib exp relates \textsc{Imp} tokens to $\uppi$ Lib Id, \textsc{Imp} arithmetic and boolean expressions to $\uppi$ Lib Exp. In particular, the compilation of an \textsc{Imp} token 
has to check if the token is a primitive type, either Rat (for Rational numbers) or Bool (for Boolean values), or an identifier. Since Rat and Bool are tokenized and we need Maude meta-level descent function downTerm to help us parse them into proper constants. 
\begin{maude}
 op compileId : Qid -> Id .
 eq compileId(I:Qid) = idn(downTerm(I:Qid, 'Qid)) .
 op compileId : Term -> Id .
 eq compileId('token[I:Qid]) =
    if (metaParse(upModule('RAT, false), 
         downTerm(I:Qid, 'Qid), 'Rat)  :: ResultPair)
    then rat(downTerm(getTerm(metaParse(upModule('RAT, false),
	      downTerm(I:Qid, 'Qid), 'Rat)), 1/2))
    else
      if (metaParse(upModule('BOOL, false), 
           downTerm(I:Qid, 'Qid), 'Bool) :: ResultPair)
      then boo(downTerm(getTerm(metaParse(upModule('BOOL, false),
		    downTerm(I:Qid, 'Qid), 'Bool)), true))
      else idn(downTerm(I:Qid, 'Qid))
      fi
    fi .
\end{maude}

To conclude this Section, the compilation of \textsc{Imp} arithmetic expressions simply maps them to their prefixed syntax counterpart in $\uppi$ Lib, e.g, an \textsc{Imp} expression a + b is compiled to add(compileExp(a), compileExp(b)).
\begin{maude}
 op compileExp : Term -> Exp .
 ceq compileExp(I:Qid) = compileId(I:Qid) if not(I:Qid :: Constant) .
 eq compileExp('token[I:Qid]) = compileId('token[I:Qid]) .
 eq compileExp('_+_[T1:Term, T2:Term]) = add(compileExp(T1:Term), compileExp(T2:Term)) .
 eq compileExp('_-_[T1:Term, T2:Term]) = sub(compileExp(T1:Term), compileExp(T2:Term)) .
 eq compileExp('_*_[T1:Term, T2:Term]) = mul(compileExp(T1:Term), compileExp(T2:Term)) .
 eq compileExp('_/_[T1:Term, T2:Term]) = div(compileExp(T1:Term), compileExp(T2:Term)) .
\end{maude}
%
%
%

\begin{figure}
\begin{center} \includegraphics[width=\columnwidth]{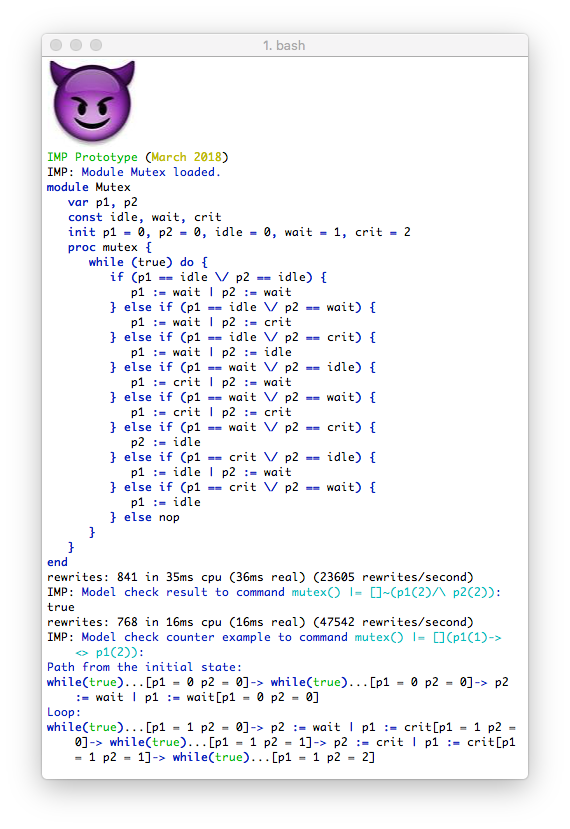} \end{center}
\caption{Loading and model checking a Mutex protocol in \textsc{Imp} that is safe but not live.}
\label{fig:imp-mutex}
\end{figure}

Figure~\ref{fig:imp-mutex} illustrates loading and model checking an \textsc{Imp} program, implementing a Mutex protocol, for safety and liveness properties.  (\textsc{Imp} command $\mid$ denotes non-deterministic choice.) State propositions $p_1$ and $p_2$ are \emph{automatically} generated by the compiler to properly construct the $\uppi$ Automata for Mutex, as discussed in Section~\ref{sec:mc-gia}. 

\section{Related work}\label{sec:rel-work}

First and foremost there is the work by Peter Mosses on Component-Based Semantics~\cite{Mosses:2008:CDP:2227536.2227559} and funcons~\cite{Churchill2015,10.1007/978-3-319-12904-4_12}, where programming language constructs are specified in Modular Structural Operational Semantics (MSOS). $\uppi$ Lib is inspired by this research and is also a result of the research on the relation between MSOS and Rewriting Logic, with an implementation in Maude, that started in~\cite{amast04,bragaMeseguer:WRLA04}, with Edward Hermann Haeusler, Peter Mosses and José Meseguer, and continued with Fabricio Chalub~\cite{Chalub:jucs_10_7:a_modular_rewriting_semantics,wrla06}. Despite their common roots, funcons and $\uppi$ Lib have different models. The models of funcons are Arrow-labeled Transition Systems and $\uppi$ Lib descriptions are to be interpreted as $\uppi$ Automata, as described in Section~\ref{sec:gia}. $\uppi$ Automata can be understood as unlabeled transition systems. This makes it easy to relate $\uppi$ Automata with term rewriting systems and to have an efficient implementation of them when transition rules are mapped to unconditional rewrite rules. This is in contrast, for instance, with previous work by Chalub and the author in the MSOS Tool in Maude~\cite{wrla06}, that understands transition rules in MSOS as conditional rewrite rules in Maude.

In~\cite{10.1007/978-3-319-12904-4_12}, Mosses and Vesely propose an implementation of Com\-po\-nent-Based Semantics using the K Framework (e.g.~\cite{rosu-serbanuta-2010-jlap}). K aims at being a methodology to define languages with tools for formal language development.
It is based on concepts from Rewriting Logic Semantics, with some intuitions from Chemical Abstract Machines~\cite{BERRY1992217} (CHAMs) and Reduction Semantics~\cite{FELLEISEN1992235} (RS). Abstract computational structures contain context needed to produce a future computation (like continuations).
Computations take place in the context of a configuration, which are hierarchically made up of K \emph{cells}. Each cell holds specific pieces of information such as computations, the environment, and memory store. 
K specifications allow for equations and rules. Equations (representing heating and cooling processes) manipulate term structure as opposed to rules that are computational and may be concurrent, similar to how Rewriting Logic understands equations and rules. K has stablished itself as a powerful framework for language semantics (e.g. the formal semantics for the Ethereum Virtual Machine~\cite{kevm}). However, it has a non-trivial model, with many different concepts, coming from different frameworks such as MSOS, Rewriting Logic, Reduction Semantics, and CHAM. 
The combination of Component-Based Semantics and K in~\cite{10.1007/978-3-319-12904-4_12} provides indeed a powerful tool for language semantics descriptions.

$\uppi$ Automata, as described in Section~\ref{sec:gia}, is a less ambitious framework while compared with K, being conceived to be simple, easily integrated into an undergraduate level course, and with an efficient implementation in Maude, as K is. As a matter of fact, it has several intersections with K given their common roots in Rewriting Logic Semantics and MSOS. Due to $\uppi$ Automata' simpler automata-based model, it appears that it is a nicer candidate to teach formal semantics and compiler construction than K. It smoothly connects with Introduction to Programming Languages, Programming Languages Semantics and Formal Languages and Automata Theory, with good properties such as expressivity, efficiency and support for automata-based automated specification and reasoning. 

\section{Conclusion}\label{sec:conclusion}

\emph{Summary.} This paper discusses the $\uppi$ framework for teaching formal compiler construction. It has a denotational character, and its implementation called $\uppi$ Lib builds on Peter Mosses Component-Based Semantics~\cite{Mosses:2004:FCF}. The framework implements a library of common programming languages constructions, such as assignments, function declarations and function calls. The semantics of a programming language is then given in a syntax-directed way, by expressing the denotations of the given programming language constructs in terms of $\uppi$ Lib elements. The semantics of $\uppi$ Lib is also described formally. Each element in $\uppi$ Lib is specified in terms of $\uppi$ Automata. Essentially, a $\uppi$ Automata describes both static and dynamic semantics by means of (unconditional) rules that relate sets of semantic components, such as the memory store, the environment, a control stack and a value stack. 
$\uppi$ Automata is overloaded to refer also to a $\uppi$ Automata with a set of state propositions that are used to validate a given $\uppi$ Automata using automata-based techniques such as model checking.
$\uppi$ Automata is a generalization of Plotkin's Interpreting Automata~\cite{plotkin}. 
%
Currently, the $\uppi$ approach is implemented in Maude yielding an effective tool for formal compiler construction and program verification. The latter is accomplished when the formal tools in Maude, such as term rewriting, narrowing and LTL model checking, are lifted to a given programming language in $\uppi$. The current prototype implementation of $\uppi$ in Maude is available at \url{http://github.com/ChristianoBraga/BPLC}, with an implementation for an imperative language called \textsc{Imp}, available in the same repository. 

\emph{Preliminary assessment and future work.} $\uppi$ appears to be a suitable approach to teach compiler construction, as much as Component-Based Semantics is to teach formal semantics of programming languages~\cite{Mosses:2004:FCF} since one only works with a small set of programming constructions that may be used to give semantics to different programming languages, in different paradigms, with a model amenable to automated verification, as discussed in Section~\ref{sec:mc-gia}. 
The approach proposed in this paper has been class tested for the past year, with quite positive results. All students have completed their projects within the academic semester, reporting it back as a rewarding experience, with a lot of work. The $\uppi$ framework appears to ease understanding of the meaning of the constructions when compared to their SOS counterparts. 
Even though a complete Maude implementation of $\uppi$ is available as reference, ways of stimulating its use and sandboxing with it need to be developed.
%
In this context, perhaps an interesting discussion regards the definition of a meta-language for describing $\uppi$ compilers. At first, our intention is to make the $\uppi$ Library available in different programming languages and let one choose one's preferred parsing/transformation framework. However, this choice appears to have some undesirable pedagogical consequences. The Maude implementation of $\uppi$, for instance, uses  meta-programming techniques that create some resistance to the understanding of the rather simple aspects of $\uppi$ Lib and its $\uppi$ Automata semantics. 
%
The author foresees the continuation of this work by addressing the issues raised in this preliminary assessment and by extending the $\uppi$ Lib library with new constructs, improving code generation and validation techniques.

\paragraph{Acknowledgements}
The author would like to \emph{warmly} thank Fabricio Chalub, Narciso Mart\'i-Oliet and Leonardo Moura for their comments on a draft of this paper, and Fabricio Chalub, Edward Hermann
Hauesler, Jos\'e Meseguer and Peter D. Mosses for the long term
collaboration that inspired the work discussed in this manuscript.


\def\sortunder#1{}

\end{document}